# Generation of high-energy clean multicolored ultrashort pulses and their application in single-shot temporal contrast measurement

Peng Wang [1], Xiong Shen [1], Jun Liu [1, 2, 3]* and Ruxin Li [1, 2, 3]*

[1]State Key Laboratory of High Field Laser Physics, Shanghai Institute of Optics and Fine Mechanics, Chinese Academy of Sciences, Shanghai 201800, China

[2]CAS Center for Excellence in Ultra-intense Laser Science (CEULS), Shanghai Institute of Optics and Fine Mechanics, Chinese Academy of Sciences, Shanghai 201800, China

[3]IFSA Collaborative Innovation Center, Shanghai Jiao Tong University, Shanghai 200237, China

*Corresponding author: jliu@siom.ac.cn, ruxinli@mail.siom.ac.cn

## ABSTRACT

We demonstrate the generation of 100-μJ-level multicolored femtosecond pulses based on a single-stage cascaded four-wave mixing (CFWM) process in a thin glass plate. The generated high-energy CFWM signals can shift the central wavelength and have well-enhanced temporal contrast because of the third-order nonlinear process. They are innovatively used as clean sampling pulses of a cross-correlator for single-shot temporal contrast measurement. With a simple home-made setup, the proof-of-principle experimental results demonstrate the single-shot cross-correlator with dynamic range of $10^{10}$, temporal resolution of about 160 fs and temporal window of 50 ps. To the best of our knowledge, this is the first demonstration in which both the dynamic range and the temporal resolution of a single-shot temporal contrast measurement are comparable to those of a commercial delay-scanning cross-correlator.

## 1．INTRODUCTION

Petawatt-level ultra-intense laser pulses have been achieved based on chirped-pulse amplification (CPA) and optical parametric chirped-pulse amplification (OPCPA) techniques [1-3], which were widely used to investigate numerous laser-matter interaction phenomena in the relativistic region in the past decades [4-11]. In some of these important research areas, such as proton electron acceleration in thin solid targets [12-14] and electron generation in fast-ignition inertial confinement fusion [15, 16], the temporal contrast of the driving laser pulse had been proved to be an extremely important parameter. For such an ultra-intense laser pulse, the prepulse and amplified spontaneous emission (ASE) noise prior to the main pulse are detrimental since their intensities might be high enough to ionize the target and thereby change its status[12-16], which makes high temporal contrast critically important for an ultra-intense laser pulse. Therefore, the characterization of temporal contrast of an ultra-intense laser pulse is the first important step towards temporal contrast improvement and precise analysis of these laser-matter interaction experiments.

In previous work, during temporal contrast measurement, the first and foremost process is the generation of a temporally clean sampling pulse [17-22]. For laser pulses with high repetition rates, a delay-scanning third-order correlator can be used to characterize the temporal pulse contrast [17, 23]. In this case, a second-harmonic generation (SHG) process using a nonlinear crystal (such as BBO) generates a clean SHG pulse that is used as the sampling pulse. Then, in another nonlinear crystal, the third-order correlation signal is obtained through the sum frequency mixing (SFM) process between the SHG sampling pulse and the test pulse and detected using a high-dynamic-range PMT. The dynamic range can be extended to $10^{12}$ by adding gradient attenuators before the detector [22, 23]. The temporal resolution can be controlled by step resolution of the delay-scanning stage and is also dependent on the group velocity mismatch between the sampling and test pulse in the crystal. However, the delay-scanning third-order correlator requires thousands of pulses to accomplish the measurement, which is time-consuming and capable of only characterizing high-repetition stable laser pulses. Besides, since the temporal contrast improvement of the SHG process is not good enough because of a second-order nonlinear process, relatively intense postpulses following the SHG sampling pulse will lead ghost prepulses into the third-order cross-correlation signal, which will affect the measurement accuracy.

As most petawatt lasers are running in single-shot mode or at low repetition,

measurement of the single-shot pulse temporal contrast is necessary. The cross-correlator is the main method for high-dynamic single-shot temporal contrast measurement, where temporal contrast information is encoded into a line of signals in space using SFM in a nonlinear crystal with a large crossing-angle [18] . So far, the highest dynamic range with a single-shot measurement was obtained by the Qian group [21]. In their optical design, a 1040 nm clean sampling pulse is generated by using a typical optical parametric amplifier (OPA) scheme so as to single-shot characterize the temporal contrast of an 800 nm ultrashort laser pulse. The OPA system consists of an optical parametric generator (OPG) stage and an OPA stage using nonlinear crystals. Since the final cross-correlator signal is collected through a fiber-array-based detection system, the temporal resolution is limited to about 0.7 ps, which is not precise enough for ultra-intense pulses with tens of femtosecond duration. The self-referenced spectral interferometry (SRSI) method was demonstrated recently by encoding the temporal contrast measurement into the frequency domain [24]. In this case, a resolution down to 20 fs can be obtained. However, the obtained best temporal contrast dynamic range has hereto been about $10^8$. Are there any methods that can simplify the generation of the clean sampling pulse? Is a new design possible to obtain high dynamic range and good temporal resolution simultaneously in a single-shot temporal contrast measurement?

We note that cascaded four-wave mixing (CFWM) has recently been studied to simultaneously generate multicolored femtosecond sidebands from UV to NIR [25-29]. Spatially and spectrally separated multicolored femtosecond sidebands are observed when two femtosecond laser pulses with different wavelengths are focused into a glass plate with a small crossing angle. As a third-order nonlinear process, the generated first-order CFWM signal has a cubic dependence on the intensity and its temporal contrast would be improved drastically, similarly to the effect of the self-diffraction (SD) and crossing polarization generation (XPW) process [30-32]. Furthermore, the temporal contrast of the first-order CFWM signal is better than that of the SHG sampling pulse which is the signal of the second-order nonlinear process. Consequently, the CFWM signal seems to be the perfect sampling pulse for a cross-correlator. However, the highest pulse energy of generated first-order CFWM signal was in a range of less than ten μJ [27] in previous work, which limits its application in a high dynamic range cross-correlator due to foreseeable low correlation signal output.

In this paper, we greatly improve the output pulse energy of first-order CFWM signals to the 100-μJ-range, which is one-order higher than previously obtained highest first-order CFWM signal. We do this in a single-stage CFWM process with a thin glass plate in the first step. It should be noted that it usually takes one stage of OPG and at least one stage of OPA to obtain 100-μJ-level wavelength-shifted amplified signals using the OPA system. A compact single-shot cross-correlator was successfully constructed with a dynamic range of about $10^{10}$, an improved temporal resolution of about 160 fs, and a temporal window of 50 ps, where the high-energy clean first-order CFWM signal is used as the sampling pulse and a 16-bit sCMOS camera is used as the correlation signal acquisition device.

## 2. EXPERIMENT

2.1 Generation of high-energy CFWM signals

Previously, the highest pulse energy obtained in the first-order CFWM signal is less than ten μJ, which is not high enough to serve as the sampling pulse for a high dynamic range temporal contrast measurement [27]. In previous work, two incident laser beams for CFWM were focused onto a glass plate using a spherical lens or concave reflective mirrors [25-29]. Self-focusing and filamentation may appear around the focal point of two high energy input beams, which will affect the generated signal. To avoid these problems and increase the input pulse energy, we use cylinder lenses to focus the two incident beams into a thin glass plate. Resultantly, pulse energies of generated CFWM signals are enhanced with increasing pulse energy of the two input beams. The proposed setup for high-energy CFWM signal generation is shown in Figure 1. After passing through a short-pass dichroic mirror with a ~800 nm cutoff, the incident ultrashort laser beam is split into two beams. Both the reflected beam with wavelength above 800 nm and the transmitted beam with wavelength shorter than 800 nm are focused by two cylindrical lenses with 500 mm focal-length onto a fused silica glass wedge with thick end of 1 mm and thin edge of 0.15 mm. A time-delay stage is used in the optical path of one of the beams to tune time delay for temporal overlapping of the two beams.

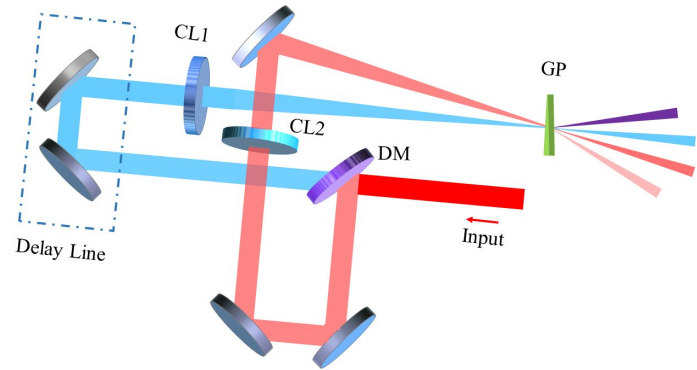

Figure 1 Schematic representation of the experimental setup for CFWM. DM is the dichroic mirror. CL1 and CL2 are two cylindrical lenses. GP is a glass plate wedge.

A proof-of-principle experiment is conducted with a Ti:sapphire femtosecond laser (Legend Elite, Coherent) delivering 35 fs, 805 nm central wavelength laser pulses at a repetition rate of 1 kHz. Limited by the output of the Ti:sapphire amplifier, only about 2.6 mJ of laser pulse power is guided into the experimental setup. After the short-pass dichroic mirror, the incident pulse energy of the long and short wavelength components are 1.6 mJ and 1.0 mJ, respectively. Bright sidebands appear on the sides of the two incident beams when they are synchronously overlapped in the glass plate in both time and space. The photograph in Figure 2(a) shows the multicolored CFWM sidebands on a white screen placed 200 mm after the glass plate. Low order CFWM signals are well separated in space, while higher-order CFWM signals are close to each other and even form a bright line due to the phase matching condition. Their spectra, measured up by a spectrometer (USB4000, Ocean Optics) cover the 660 nm to 890 nm range and are smoother than those of the input beams as shown in Figure 2(b).

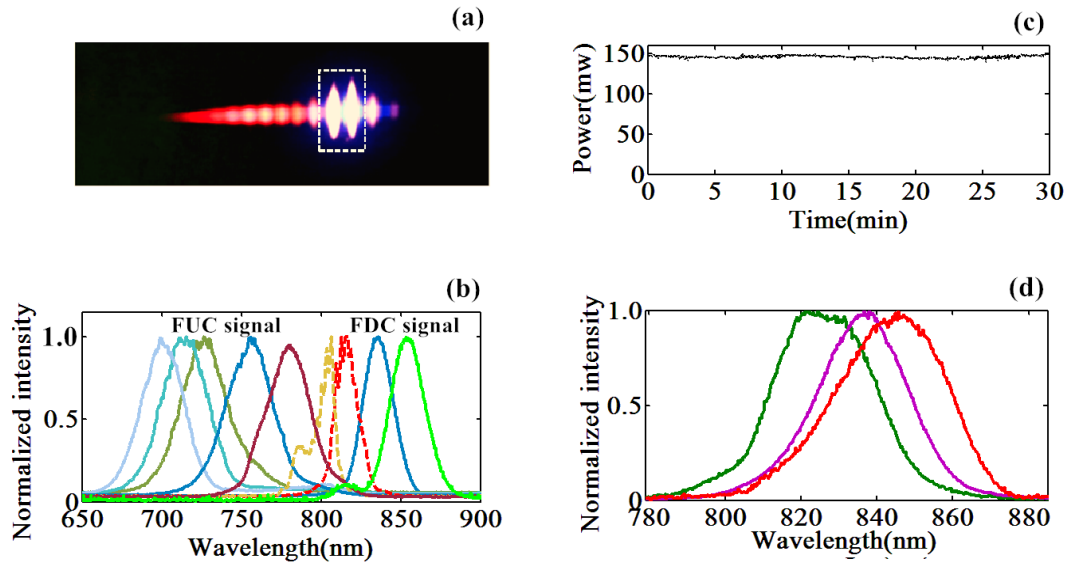

Figure 2 The generation of CFWM signals. (a) Photograph of the generated signals. The spots in the dashed rectangle are input beams. (b) Spectra of the generated signals (solid lines) and the input beams (dashed lines). (c) Power stability of first-order FDC signal. (d) Central wavelength tunability of the first-order FDC signal.

When the crossing-angle of the two input beams is 2° in the air, the pulse energies of the first-order frequency-upconversion (FUC) and frequency-downconversion (FDC) CFWM signals are 60 μJ and 146 μJ, respectively. The power stability of the first-order FDC signals is monitored using a power meter (PM100D, Thorlabs) for half an hour resulting in 0.97% RMS power stability, as shown in Figure 2(c), while that of the incident laser is about 0.46% RMS. The central wavelength of the generated first-order FDC signal can be conveniently tuned by changing the delay time between the two incident beams as shown in the Figure 2(d). As in our previous work [30], the beam quality achieved is satisfactory (data not shown in this paper). The output power can be further increased and the tunable spectral range of the first-order signal can be extended if a 25 fs Ti: sapphire laser system with higher pulse energy and broader spectrum is used. Furthermore, the power stability can further be improved if external experimental conditions such as air-flow and vibration in the lab are better controlled.

Frequency-degenerated four-wave mixing processes such as XPW and SD can be used to improve the temporal contrast of ultrashort laser pulses [30-32]. In the same manner, the CFWM process is a frequency non-degenerated cascaded four-wave mixing process, where the generated CFWM signals provide an improved temporal contrast in comparison to that of the input laser pulses. To verify pulse cleaning ability of the CFWM process, the temporal contrast of both the input pulse and the first-order FDC signal are measured using a commercial third-order cross-correlator (Amplitude Technologies Inc., Sequoia 800). Results are shown in Figure 3. Obviously, the first-order FDC signal is sufficiently cleaned with a temporal contrast improvement of about $10^6$ in comparison to that of the input laser pulse. It should be noted that since the Sequoia is optimized for 800 nm central wavelength, $10^{10}$ temporal contrast already marks the measurement limit for the first-order FDC signal with 146 μJ pulse energy. As can be deduced from above results, the first-order CFWM pulse holds both a 100-μJ-level pulse energy and high temporal contrast, which means that it can be used as the sampling pulse in a single-shot cross-correlator.

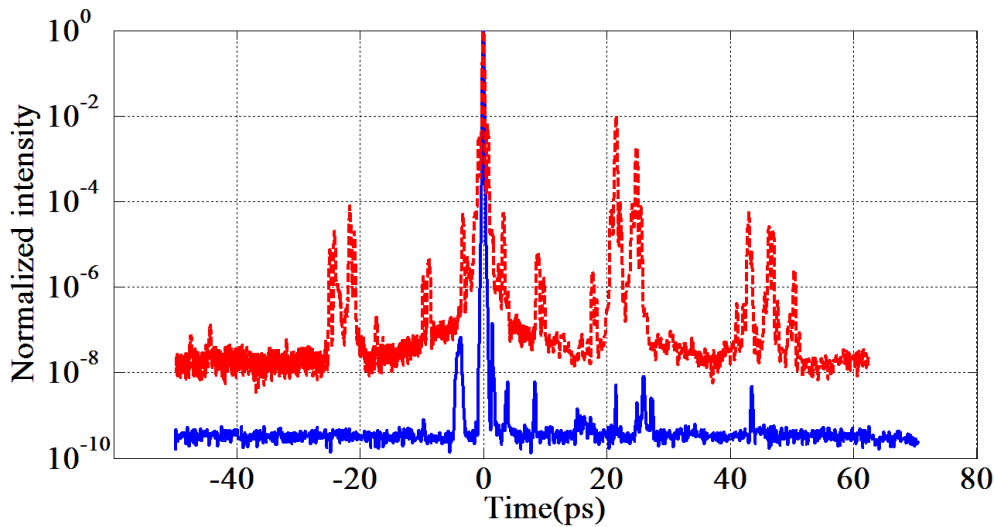

Figure 3 Temporal contrast curves of the input pulse (red dashed line) and the first-order FDC signal (blue solid line).

2.2 Single-shot temporal contrast measurement

The third-order correlator has been the dominant method for single-shot temporal contrast characterization in the past decades since its first application in 2001[18]. The dynamic range, temporal resolution and temporal window are three key parameters for a single-shot temporal contrast measurement. The single-shot measurement is based on the time-to-space encoding in which the temporal intensity measurement is transformed to a spatial intensity measurement with a spatial detector as shown in the Figure 4. Consequently, in order to improve the dynamic range of a temporal contrast measurement, a simple method is to obtain the maximal correlation signal and then introduce an attenuator to the strongest main pulse. So far, the highest single-shot dynamic range is $10^{10}$ where a point attenuator is used to weaken the main pulse by about four orders of magnitude [21]. To improve the temporal window, a larger size of nonlinear crystal and a larger beam diameter with a relatively larger crossing-angle are typically used during the final SFM process. The best result to date is about 50 ps [21]. Several other methods have been applied to further extend the time window. For example, a pulse replicator was used to extend the temporal window to about 200 ps [20]. As for the temporal resolution of a single-shot third-order cross-correlator, there are three limitations. The first limitation comes from the group velocity mismatch (GVM) in the nonlinear crystal. The GVM can be expressed as $\tau = l * (\frac{1}{v_s cos\theta} - \frac{1}{v_p cos\alpha})$, where $l$ is the length of the SFM nonlinear crystal, and $v_s$ and $v_p$ are group velocities of the sampling pulse and test pulse, respectively. Angles $\theta$ and $\alpha$ denote crossing-angles of the test pulse and the sampling pulse, respectively, with respect to the propagation direction of correlator signal in the nonlinear crystal. The second limitation arises from the pixel size of the cross-correlation signal detector. Finally, the third limitation is the spatial resolution of the 4f imaging system, which is used to map the correlation signal from the nonlinear crystal to the detection system. To date, temporal resolutions for single-shot cross-correlators are in the range of hundreds of femtoseconds [21]. When all three key parameters are taken into consideration, the best results of single-shot cross-correlators to date was obtained by the Qian group in 2014 with a fiber array detection system, where one OPG and one stage of OPA were used to generate the sampling pulse [21].

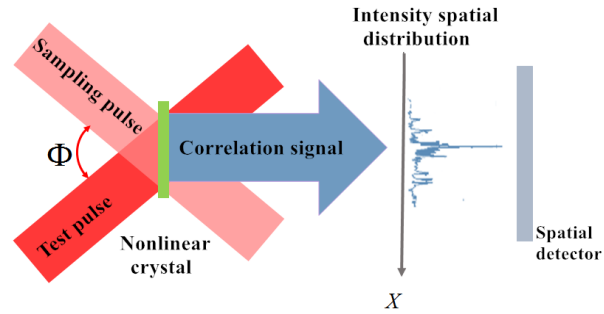

Figure 4  The time-to-space encoding for a single-shot cross-correlator.

In this study, employing advantages from the setup of the Qian group [21], we construct an improved and simplified cross-correlator for single-shot temporal contrast measurement by using the first-order CFWM signal as the sampling pulse. Figure 5 illustrates the optical schematic setup of the novel cross-correlator. The femtosecond laser input pulse is split into two by using a beam splitter with a 20:80 reflection/transmission (R:T) ratio. The reflected signal is applied as the test pulse. The stronger transmitted signal is used as the input pulse for high-energy CFWM signal generation in a thin glass plate wedge. The generated first-order CFWM signal and the test pulse are firstly filtered by two soft-edge irises to obtain an almost top-hat intensity distribution, and then are spatially expanded by two separate beam expanders (GBE02-B, THORLAB), to be focused into a wedge-designed nonlinear BBO crystal by using two 200 mm focal-length concave cylindrical reflective mirrors with a large crossing-angle. The cross-correlation signal is produced by the SFG process in the BBO crystal and imaged onto a 16-bit sCMOS camera by using a 4f lens pair. In order to achieve a high dynamic range measurement, we also attenuate the main peak of the cross-correlation signal by using a strip-shaped density filter, which leads to about 4 orders of magnitude attenuation. Furthermore, a coated wedge is placed right behind the BBO to introduce attenuated signals for the generation of a reference replica signal below the main cross-correlation signal, whose attenuated main pulse still saturates the camera. Moreover, to avoid scattering noise from the test and sampling pulses, a short-pass filter with a cutoff at 700 nm is placed in front of the sCMOS camera.

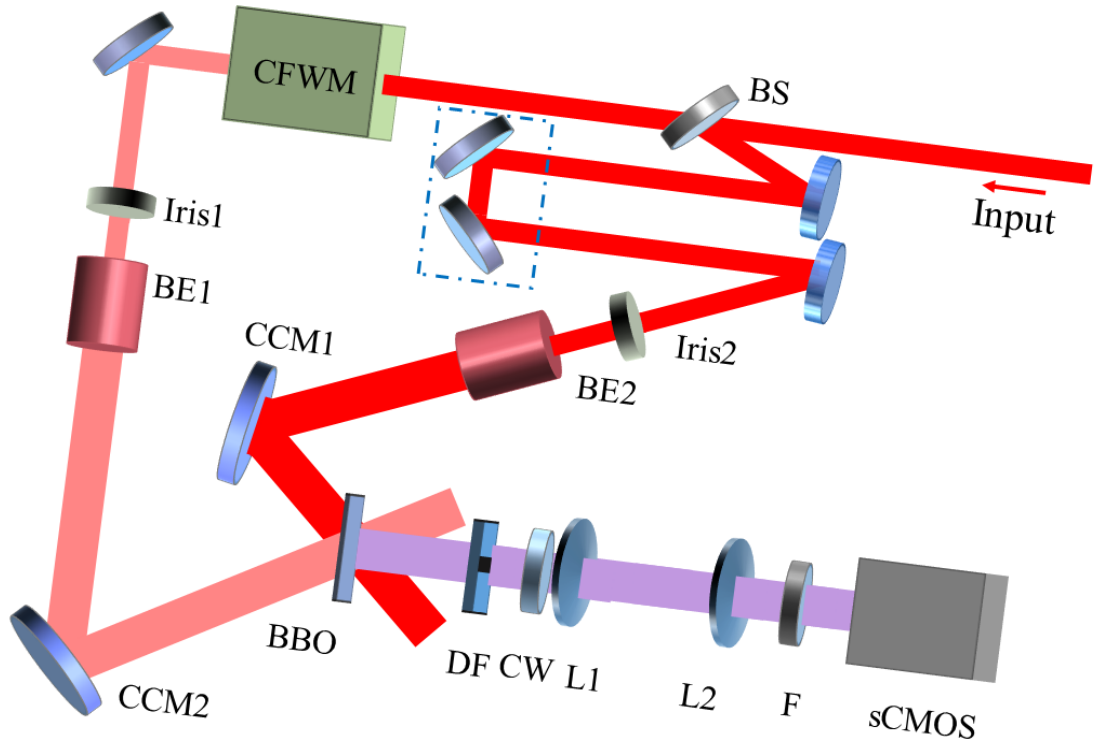

Figure 5 Schematic representation of the experimental setup in single-shot temporal contrast measurement. BS denotes the beam splitter. BE1 and BE2 are beam expanders. CCM1 and CCM2 are concave cylindrical mirrors. DF is a strip-shaped density filter. CW is a coated wedge. L1 and L2 are 4f imaging lenses. F is a short-pass filter with a cutoff at 700 nm. Iris1 and Iris2 are two home-made soft-edge irises.

As a proof-of-principle experiment, the femtosecond Ti: sapphire regenerative amplifier (1 kHz/40 fs) is used to test the novel single-shot temporal contrast method. About 2.6 mJ laser pulses are guided into the experimental setup. Since approximately 0.6 mJ laser power is used in the test pulse, only about 2.0 mJ power is used for the CFWM process, and therefore about 100 μJ first-order FDC signal is generated. A 15 mm × 15 mm × 1.5 mm BBO crystal is cut at an angle of 80° to ensure that the phase-matching angle between the sampling and test pulse is about 63° in the air, which helps expand the temporal window to about 50 ps.

According to the principle of a third-order cross-correlator, the temporal intensity of the test pulse is mapped onto a 1-D spatial-intensity distribution based on the time-to-space encoding. The correlation signal intensity is highly dependent on the intensities of the two incident pulses, and the uniformity of intensities across the interaction region is important for dynamic range accuracy. Two processes are applied to improve this accuracy. The first one involves obtaining relatively uniform distribution intensity of input beams from the Gaussian beam. Central parts of input beams are firstly filtered by the soft edge irises and then expanded to 13 mm diameter by two beam expanders to fit the size of the BBO crystal. The second process involves calibration measurement. Primarily, when the sampling and test pulse are synchronously overlapped in the middle of the interaction region, relative time delay is set as 0 ps. By changing time delay between the two pulses, they are synchronously overlapped in time at different positions in the interaction region, and the intensity of the peak of the cross-correlation signal is captured by the sCMOS as shown in the Figure 6. The fitted curve based on calibration data at different positions is added to the final data analysis to improve dynamic range accuracy. It needs to be noted that this spatial uniformity and beam expansion, as well as the calibration procedure, is unnecessary during the measurement of a petawatt laser pulse, since it is usually a top-hat beam with large enough beam size.

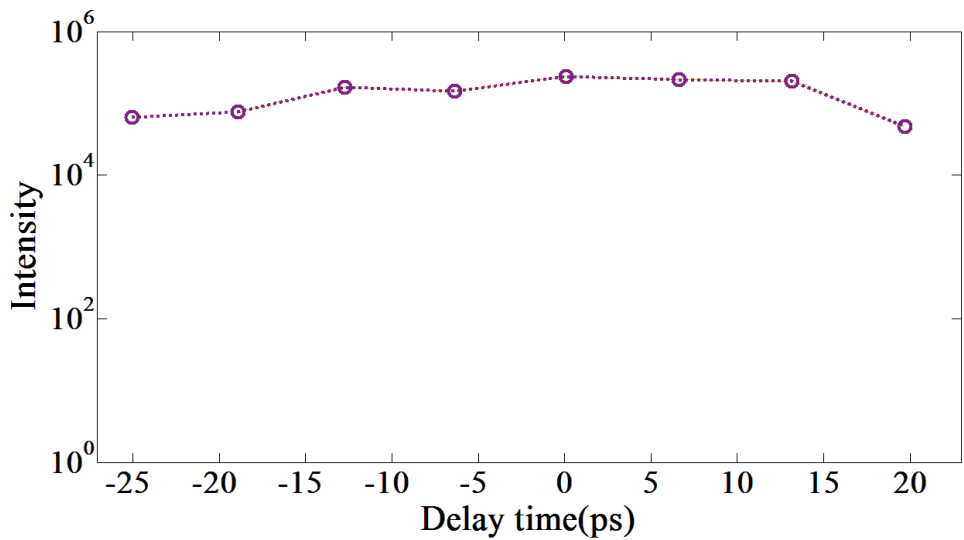

Figure 6 Intensity of the main pulse of the correlator signal relative to the time delay between the sampling and test pulses.

In the equipment of this study, a strip-shaped density filter with an attenuation of 10 000× is placed right behind the BBO crystal to weaken the strongest main signal. Even after passing through the density filter, the main pulses are still too powerful for the detector, while the addition of another density filter would make the main pulse too weak to be detected. Therefore, we use a wedge glass plate, which can provide a replica-correlation signal with an intensity attenuation of about 70×. The replica, located at a different region on the sCMOS chip, helps calibrate the intensity of the prepulse, postpulse and the attenuated main pulse, which still saturate the sCMOS detector. The wedge is not always necessary, as in the case where the peak signal after the strip-shaped density filter does not saturate the detector. Finally, the preprocessed cross-correlation signal is imaged onto a 2048×800 pixel area of the sCOMS and captured with an exposure time of 1 ms which is a single-shot measurement for 1 kHz repetition laser pulses. An example of the single-shot result captured by the sCMOS is shown in the Figure 7 in which Ⅰ, Ⅱ, and Ⅲ are the main pulse after passing through the strip-shaped density filter and two prepulses that saturate the sCMOS detector, respectively. Meanwhile, 1, 2 and 3 denote replicas of Ⅰ, Ⅱ, and Ⅲ provided by the coated wedge, respectively.

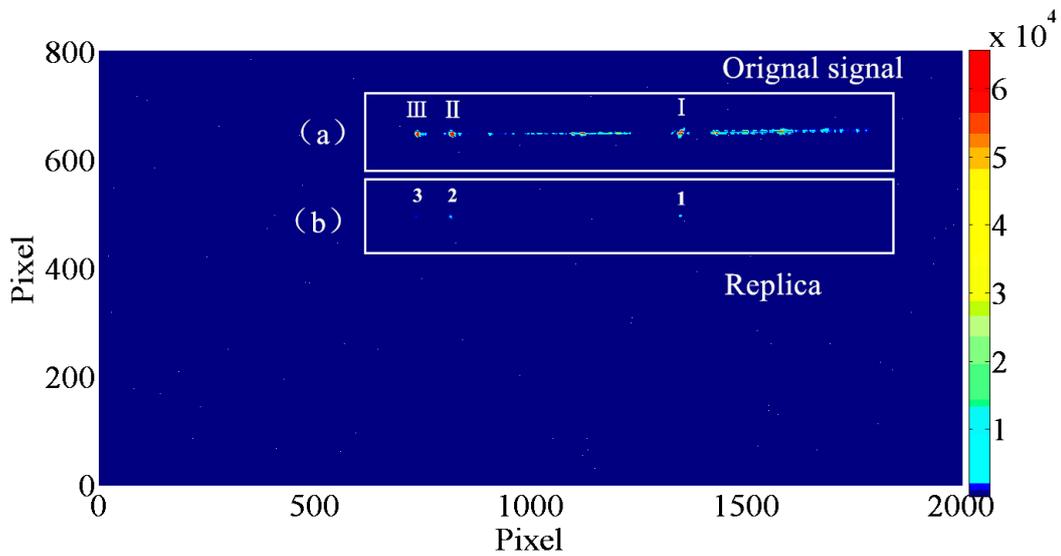

Figure 7 Original single-shot and replica correlator signals on the sCMOS detector. (a) Original signal, where the main pulse (I) is attenuated by a strip-shaped density filter. (b) Replica of the original signal provided by a coated wedge.

By summing up the intensity of pixels along the vertical direction and subtracting the background noise, we obtain the temporal contrast of the test pulse within the measurement temporal window. And for a single-shot measurement, the temporal window is about 50 ps, limited by the aperture of the nonlinear crystal, the width of the input beam, and the crossing-angle. By changing the delay-time between the sampling pulse and test pulse by about 40 ps, we performed another shot of measurement. Thus, the total integrated temporal window can be extended to about 90 ps as shown in Figure 8, where both shots of measurement has a 10 ps overlap region to help joint the results. And so on, we can obtain about 130 ps temporal window with three shots of delay-shifted measurement. Actually, since only about 2/3 part of the detector is used by the correlation signal, we expect to obtain wider single-shot temporal window in the near future with larger beam diameter and wider nonlinear crystal.

The noise aroused from SHG signals of the sampling pulse and test pulse need to be analyzed. By blocking either the sampling pulse or test pulse, the intensity of noise is captured by the sCMOS camera. This signal is compared with the background of the sCMOS as shown in the bottom of Figure 8 as well. Because of the large phase mismatch and clean BBO crystal surface, the noise scattered from both SHG signals are comparable to the background of the sCMOS camera. A clean surface of the nonlinear crystal is very important, since any dust or flaws would cause severe scattering of light. According to SHG noise, a dynamic range as high as $3\times10^{10}$ can be achieved by using current setup. An even higher dynamic range is expected to be attainable by using a higher energy sampling pulse together with a detector of relatively higher sensitivity and dynamic range. Besides, band pass filter centered at around 409 nm would help to void the further influence of the noise from the SHG signals because of the spectrum difference of the sampling pulse and test pulse. Temporal contrast of the test pulse is also measured by using a commercial delay-scanning third-order correlator (Amplitude Technologies Inc., Sequoia-800) as a reference for comparison, as shown in Figure 8. For verification of the accuracy of the measurement, a 4 mm thickness glass plate is used to introduce a postpulse for the test pulse. The beam splitter and inserted 4 mm thickness glass plate introduce postpulses C and D, respectively. The two measurements show perfect agreement with each other within the given temporal window, which speaks high reliability and capability of our scheme.

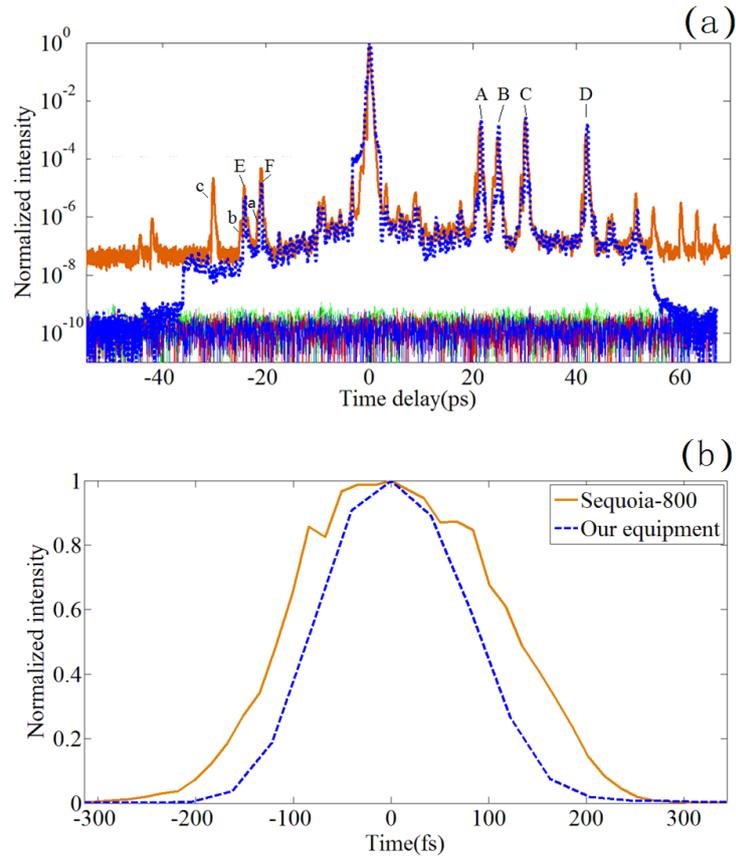

Figure 8 (a) Temporal contrast results of a Ti: sapphire regenerative amplifier measured by using the Sequoia (the orange line) and our single-shot equipment with two shots of measurement (the blue dotted line). At the bottom of the graph, three lines are presented to show the magnitude of noise from SHG signals. The red line depicts the result when the test pulse is blocked. The green line is the result when the sampling pulse is blocked. The solid blue line is the background signal (when both are blocked) of the sCMOS. (b) Correlation traces by the Sequoia and our equipment with a linear plot of intensity (or enlargement of the main pulse).

In more detail, we find that even though the prepulses E and F are detected by both Sequoia and our equipment, there are three prepulses marked as c, b and a with the intensity ratios of about $2.2 \times 10^{-5}$, $0.7 \times 10^{-6}$, and $1.7 \times 10^{-6}$, located at around t = –30.21 ps, t = –24.91 ps, and t = –21.49 ps, respectively, that are only detected by the Sequoia-800. The prepulses a and b are very close to the prepulses E and F. Indeed, this difference between Sequoia-800 and our single-shot device proves another advantage of our equipment. This is because the measured prepulses a, b and c are actually ghost pulses introduced by the postpulse of the sampling SHG pulse when Sequoia-800 is used. Postpulses of the test pulse, marked as A, B and C, with the intensity ratios of about $1.3 \times 10^{-3}$, $0.5 \times 10^{-3}$, and $2.5 \times 10^{-3}$ and located at around t = 21.40 ps, t = 24.80 ps, and t = 30.14 ps, respectively, arise because of the back and forth reflection when the main pulse passes through some optical components. For the sampling pulse generated by the second-order nonlinear process, postpulses with intensity ratios of $10^{-6}$ or $10^{-5}$ still exist and eventually lead to the ghost prepulse with the same intensity ratio through the correlation process. In comparison, the sampling pulse in our equipment is generated by the third-order process, and its intensity can be expressed in the time domain as: $I_{FD} = I_1 * I_2^2$ where FD, 1, 2 refer to the first-order FDC signal and the two input beams for CFWM, respectively. As a result, even if the pulses have postpulses with the intensity of $10^{-3}$, the intensity of the postpulses of the sampling pulse would in theory be weakened to about $10^{-9}$ magnitude. Therefore, ghost prepulses will also be at about a $10^{-9}$ level, which is much lower than the temporal contrast of about $10^8$ in this experiment, and renders them undetectable. As a result, the high energy CFWM signal is a better sampling pulse with cleaner pulse trailing edge in comparison to the SHG signal.

Finally, we study the temporal resolution of our equipment. A femtosecond pulse (about 40 fs) from a Ti: sapphire regenerative amplifier [21] is measured using our experiment setup and the Sequoia-800. The result implies about 160 fs temporal resolution with our equipment, which is the FWHM of the measured correlation trace width. The resolution of Sequoia-800 is about 250 fs, which is consistent with the result in reference [21]. The measured correlation traces in a linear plot are shown in Figure 8(b). It is obvious that our measurement curve is narrower than that that of Sequoia-800. Therefore, it can be concluded that our equipment has a better temporal resolution than that of the commercial third-order correlator Sequoia-800. In our measurement, the 800 nm test pulse and the sampling pulse have angles with respect to the propagation direction of the correlator signal in the SFM crystal of approximately 18° and 19°, respectively, and the thickness of the crystal is 1.5 mm. Thus, the temporal resolution limited by the GVM is about 42 fs based on the equation $\tau = l * (\frac{1}{v_s cos\theta} - \frac{1}{v_p cos\alpha})$. Furthermore, we use a sCMOS with pixel size of only 11 μm × 11 μm to analyze the correlation signal. The spatial transversely distributed signal on the sCMOS is sampled by about 1261 pixels in the time window of 50 ps, resulting in the limitation on temporal resolution of the data acquisition device at about 40 fs. However, the focal length (*f*) of the 4f imaging system in our equipment is 100 mm and the width (*d*) of the main pulse of correlation signal on the lens is about 1.5 mm, which results in the width (*w*) of the Airy diffraction pattern of about 33 μm based on the equation $w = 1.22 \lambda f / d$. Three pixels of the sCMOS detector are needed to capture the signal. Thus, the theoretical temporal resolution limited by the imaging system is about 120 fs, which is close to measurement results. If the correlation signal could be mapped onto the sCMOS detector with several-times magnification, the resolution of the measurement could be further improved.

## 3. CONCLUSION

As high as 100-μJ-level powers of the first-order CFWM sidebands with shifted central wavelength and high temporal contrast are generated in a thin glass plate based on the CFWM process, which usually needs at least one stage of OPG and one stage of OPA when an OPA system is used. The first-order sideband is used as the sampling pulse in a compact and economical single-shot cross-correlator for temporal contrast measurement. The proof-of-principle experiment results have demonstrated that the single-shot correlator has a dynamic range of about $10^{10}$, a temporal resolution of about 160 fs, and a temporal window of 50 ps. To the best of our knowledge, it is the first experiment demonstrating that a single-shot cross-correlator is comparable to the commercial delay-scanning cross-correlator in terms of both the dynamic range and the temporal resolution. Furthermore, higher dynamic ranges can be obtained by increasing the correlation signal energy and sensitivity of the detector, and higher temporal resolution can be obtained by using a magnification 4f mapping setup, and wider temporal window can also be obtained by using larger input beam diameter and wider nonlinear crystals in the future.

## FUNDING

This work is supported by the National Natural Science Foundation of China (NSFC) (61521093, 61527821); the Instrument Developing Project (YZ201538) and the Strategic Priority Research Program (XDB16) of the Chinese Academy of Sciences (CAS).

## ACKNOWLEDGMENTS

The authors would like to thank Prof. Liejia Qian and Dr. Jingui Ma from Shanghai Jiao Tong University for helpful discussions.

## REFERENCES

1. D. Strickland, and G. Mourou, " Compression of amplified chirped optical pulses," Optics Communications **56**, 219-221 (1985).

2. A. Dubietis, G. Jonusauskas, and A. Piskarskas, " Powerful femtosecond pulse generation by chirped and stretched pulse parametric amplification in BBO crystal," Optics Communications **88**, 437-440 (1992).

3. C. Danson, D. Hillier, N. Hopps, and D. Neely, "Petawatt class lasers worldwide," High Power Laser Science and Engineering **3**, e3 (2015).

4. R. Betti, and O. A. Hurricane, "Inertial-confinement fusion with lasers," Nature Physics **12**, 435-448 (2016).

5. A. Macchi, M. Borghesi, and M. Passoni, "Ion acceleration by superintense laser-plasma interaction," Reviews of Modern Physics **85**, 751-793 (2013).

6. B. Dromey, M. Zepf, A. Gopal, K. Lancaster, M. S. Wei, K. Krushelnick, M. Tatarakis, N. Vakakis, S. Moustaizis, R. Kodama, M. Tampo, C. Stoeckl, R. Clarke, H. Habara, D. Neely, S. Karsch, and P. Norreys, "High harmonic generation in the relativistic limit," Nature Physics **2**, 456-459 (2006).

7. H. Daido, M. Nishiuchi, and A. S. Pirozhkov, "Review of laser-driven ion sources and their applications," Reports on Progress in Physics **75**, 71 (2012).

8. L. Robson, P. T. Simpson, R. J. Clarke, K. W. D. Ledingham, F. Lindau, O. Lundh, T. McCanny, P. Mora, D. Neely, C. G. Wahlstrom, M. Zepf, and P. McKenna, "Scaling of proton acceleration driven by petawatt-laser-plasma interactions," Nature Physics **3**, 58-62 (2007).

9. H. T. Kim, K. H. Pae, H. J. Cha, I. J. Kim, T. J. Yu, J. H. Sung, S. K. Lee, T. M. Jeong, and J. Lee, "Enhancement of electron energy to the multi-gev regime by a dual-stage laser-wakefield accelerator pumped by petawatt laser pulses," Physical Review Letters **111**, 5 (2013).

10. I. Pomerantz, E. McCary, A. R. Meadows, A. Arefiev, A. C. Bernstein, C. Chester, J. Cortez, M. E. Donovan, G. Dyer, E. W. Gaul, D. Hamilton, D. Kuk, A. C. Lestrade, C. Wang, T. Ditmire, and B. M. Hegelich, "Ultrashort pulsed neutron source," Physical Review Letters **113**, 6 (2014).

11. E. Liang, T. Clarke, A. Henderson, W. Fu, W. Lo, D. Taylor, P. Chaguine, S. Zhou, Y. Hua, X. Cen, X. Wang, J. Kao, H. Hasson, G. Dyer, K. Serratto, N. Riley, M. Donovan, and T. Ditmire, "High e + /e- ratio dense pair creation with $10^{21}$W.cm$^{-2}$ laser irradiating solid targets," Scientific Reports **5**, 12 (2015).

12. M. Kaluza, J. Schreiber, M. I. K. Santala, G. D. Tsakiris, K. Eidmann, J. Meyer-ter-Vehn, and K. J. Witte, "Influence of the laser prepulse on proton acceleration in thin-foil experiments," Physical Review Letters **93**, 045003 (2004).

13. O. Lundh, F. Lindau, A. Persson, C. G. Wahlström, P. McKenna, and D. Batani, "Influence of shock waves on laser-driven proton acceleration," Physical Review E **76**, 026404 (2007).

14. D. Batani, R. Jafer, M. Veltcheva, R. Dezulian, O. Lundh, F. Lindau, A. Persson, K. Osvay, C. G. Wahlström, D. C. Carroll, P. McKenna, A. Flacco, and V. Malka, "Effects of laser prepulses on laser-induced proton generation," New Journal of Physics **12**, 045018 (2010).

15. H.-b. Cai, K. Mima, A. Sunahara, T. Johzaki, H. Nagatomo, S.-p. Zhu, and X. T. He, "Prepulse effects on the generation of high energy electrons in fast ignition scheme," Physics of Plasmas **17**, 023106 (2010).

16. A. G. MacPhee, L. Divol, A. J. Kemp, K. U. Akli, F. N. Beg, C. D. Chen, H. Chen, D. S. Hey, R. J. Fedosejevs, R. R. Freeman, M. Henesian, M. H. Key, S. Le Pape, A. Link, T. Ma, A. J. Mackinnon, V. M. Ovchinnikov, P. K. Patel, T. W. Phillips, R. B.

Stephens, M. Tabak, R. Town, Y. Y. Tsui, L. D. Van Woerkom, M. S. Wei, and S. C. Wilks, "Limitation on prepulse level for cone-guided fast-ignition inertial confinement fusion," Physical Review Letters **104** (2010).

17. S. Luan, M. Hutchinson, R. Smith, and F. Zhou, "High dynamic range third-order correlation measurement of picosecond laser pulse shapes," Measurement Science and Technology **4**, 1426 (1993).

18. J. Collier, C. Hernandez-Gomez, R. Allott, C. Danson, and A. Hall, "A single-shot third-order autocorrelator for pulse contrast and pulse shape measurements," Laser and Particle Beams **19**, 231-235 (2001).

19. E. J. Divall, and I. N. Ross, "High dynamic range contrast measurements by use of an optical parametric amplifier correlator," Optics Letters **29**, 2273-2275 (2004).

20. C. Dorrer, J. Bromage, and J. D. Zuegel, "High-dynamic-range single-shot cross-correlator based on an optical pulse replicator," Optics Express **16**, 13534-13544 (2008).

21. Y. Wang, J. Ma, J. Wang, P. Yuan, G. Xie, X. Ge, F. Liu, X. Yuan, H. Zhu, and L. Qian, "Single-shot measurement of $> 10^{10}$ pulse contrast for ultra-high peak-power lasers," Scientific Reports **4**, 3818 (2014).

22. A. Kon, M. Nishiuchi, H. Kiriyama, K. Ogura, M. Mori, H. Sakaki, M. Kando, and K. Kondo, "High dynamic range multi-channel cross-correlator for single-shot temporal contrast measurement," Journal of Physics: Conference Series **717**, 012103 (2016).

23. H. Kiriyama, T. Shimomura, H. Sasao, Y. Nakai, M. Tanoue, S. Kondo, S. Kanazawa, A. S. Pirozhkov, M. Mori, Y. Fukuda, M. Nishiuchi, M. Kando, S. V. Bulanov, K. Nagashima, M. Yamagiwa, K. Kondo, A. Sugiyama, P. R. Bolton, T. Tajima, and N. Miyanaga, "Temporal contrast enhancement of petawatt-class laser pulses," Optics Letters **37**, 3363-3365 (2012).

24. T. Oksenhendler, P. Bizouard, O. Albert, S. Bock, and U. Schramm, "High dynamic, high resolution and wide range single shot temporal pulse contrast measurement," Optics Express **25**, 12588-12600 (2017).

25. J. Liu, and T. Kobayashi, "Generation of µJ multicolor femtosecond laser pulses using cascaded four-wave mixing," Optics Express **17**, 4984-4990 (2009).

26. W. Liu, L. Zhu, L. Wang, and C. Fang, "Cascaded four-wave mixing for broadband tunable laser sideband generation," Optics Letters **38**, 1772-1774 (2013).

27. P. Wang, J. Liu, F. Li, X. Shen, and R. Li, "Generation of high-energy tunable multicolored femtosecond sidebands directly after a Ti:sapphire femtosecond laser," Applied Physics Letters **105**, 201901 (2014).

28. P. Wang, J. Liu, F. Li, X. Shen, and R. Li, "Multicolored sideband generation based on cascaded four-wave mixing with the assistance of spectral broadening in multiple thin plates," Photonics Research **3**, 210 (2015).

29. J. Liu, and T. Kobayashi, "Generation and amplification of tunable multicolored femtosecond laser pulses by using cascaded four-wave mixing in transparent bulk media," Sensors **10**, 4296-4341 (2010).

30. J. Liu, K. Okamura, Y. Kida, and T. Kobayashi, "Temporal contrast enhancement of femtosecond pulses by a self-diffraction process in a bulk Kerr medium," Optics Express **18**, 22245-22254 (2010).

31. X. Shen, P. Wang, J. Liu, and R. Li, "Linear angular dispersion compensation of cleaned self-diffraction light with a single prism," High Power Laser Science and Engineering **6**, e23 (2018).

32. A. Jullien, O. Albert, F. Burgy, G. Hamoniaux, L. P. Rousseau, J. P. Chambaret, F. Auge-Rochereau, G. Cheriaux, J. Etchepare, N. Minkovski, and S. M. Saltiel, "$10^{-10}$ temporal contrast for femtosecond ultraintense lasers by cross-polarized wave generation," Optics Letters **30**, 920-922 (2005).